\documentclass[a4paper,11pt]{article}
\usepackage[utf8x]{inputenc}
\usepackage{amsmath,amssymb,cite,dsfont,mathtools,tensor}
\usepackage{hyperref,verbatim}

\DeclareMathOperator{\Tr}{Tr}
\numberwithin{equation}{section}

\title{{\bf A Matrix Model for QCD: QCD Colour is Mixed}}

\author{A. P. Balachandran$^{a,b}$\footnote{balachandran38@gmail.com},\,, Amilcar  de Queiroz$^{c,d}$\footnote{amilcarq@unb.br}\, and Sachindeo Vaidya$^b$\footnote{vaidya@cts.iisc.ernet.in} \\
\begin{small}{\it $^a$Department of Physics, Syracuse University, Syracuse, N. Y. 13244-1130, USA}
\end{small}\\
\begin{small}{\it $^b$Centre for High Energy Physics, Indian Institute of Science, Bangalore, 560012, India}
\end{small}\\
\begin{small}{\it $^c$Instituto de Fisica, Universidade de Brasilia, Caixa Postal 04455, 70919-970, Brasilia, DF, Brazil} 
\end{small}\\
\begin{small}{\it $^d$Departamento de F\'isica Te\'orica, Facultad de Ciencias, Universidad de Zaragoza, 50009 Zaragoza, Spain} \end{small}}

\date{\today}

\begin{document}

\maketitle

\begin{abstract}
We use general arguments to show that coloured QCD states when restricted to gauge invariant local observables are mixed. This result has 
important implications for confinement: a pure colourless state can never evolve into two coloured states by unitary evolution. Furthermore, the 
mean energy in such a mixed coloured state is infinite. Our arguments are confirmed in a matrix model for QCD that we have developed using the 
work of Narasimhan and Ramadas \cite{Narasimhan:1979kf} and Singer \cite{Singer:1978dk}. This model, a $(0+1)$-dimensional quantum 
mechanical model  for gluons free of divergences and capturing important topological aspects of QCD, is adapted to analytical and numerical work. It 
is also suitable to work on large $N$ QCD. As applications, we show that the gluon spectrum is gapped and also estimate some low-lying levels for 
$N=2$ and 3 (colors). 

Incidentally the considerations here are generic and apply to any non-abelian gauge theory.
\end{abstract}

\section{Introduction}

The understanding of physical states in QCD is of fundamental importance. Conjectures regarding quark confinement and chiral symmetry breaking 
are based on speculations about their nature. It is also important for a non-perturbative formulation of QCD.

Gribov \cite{Gribov:1977mi} showed many years ago that the Coulomb gauge in QCD does not fully fix the gauge and is inadequate for 
a non-perturbative formulation of QCD. Later, Singer \cite{Singer:1978dk} and Narasimhan and Ramadas \cite{Narasimhan:1979kf} proved that 
the Gribov problem cannot be resolved by choosing another gauge condition since 
the gauge bundle on the QCD configuration space is twisted.

In this paper, we argue that as a consequence of the above twisted nature of the QCD bundle, coloured states restricted to the algebra of local 
observables are necessarily mixed: they carry entropy. This argument is confirmed in a matrix model for gluons we also propose here. This model 
is $0+1$ dimensional and free of the technical problems of quantum field theory. 

The matrix model, being a quantum mechanical model of $8\times 8$ real matrices, for $N=3$ colours, and capturing certain essential topological aspects of QCD offers 
a new approach to QCD calculations. It is also suitable for the study of 't Hooft's large $N$ limit. As an explicit illustration of the power of our 
approach, we show that the gluon spectrum has a gap in our model. In lattice calculations this is taken as a signal for confinement.

For $N=2$ and 3, we also use simple variational calculations to estimate low-lying glueball masses. Detailed numerical work is on progress.

Just as in a soliton model, it is necessary to quantise the excitations around our matrix model solutions in a full quantum field theory. In this 
connection, we note that the matrix model contains the vacuum sector where the gluon potential is gauge equivalent to the zero field. We also 
indicate how to construct multiparticle levels for our gluon levels adapting standard techniques in soliton physics \cite{Balachandran:1991zj}.

In a paper under preparation, we will argue that QCD has different phases, and also calculate the glueball spectrum in these phases. The 
Dirac operator in the matrix model approach will also be discussed.

\section{The Gauge Bundle in QCD}

Let $A_i=A_i^\alpha (\lambda_\alpha/2)$, with $i=1,2,3$ and $\lambda_\alpha$ being Gell-Mann matrices, denote 
the QCD vector potentials (in our convention, $D_\mu = \partial_\mu + A_\mu$, with $A^\dagger_\mu = -A_\mu$) in the temporal $A_0=0$ gauge. 
Its gluon configuration space $Q$ is based on the space $\mathcal{A}=\{A=(A_1,A_2,A_3)\}$ of their connections. The QCD gauge group 
$\mathcal{G}$ is the group $\{u\}$ of maps from $\mathds{R}^3$ to $SU(3)$ with the asymptotic condition (time argument is suppressed)
\begin{equation}
  u(\vec{x})~\stackrel{|\vec{x}|\to\infty}{\longrightarrow}~u_\infty \in SU(3).
\end{equation}
(See also the Sky group in this respect \cite{Balachandran:2013wsa}.) The group $\mathcal{G}$ acts on $\mathcal{A}$ according to
\begin{equation}
  u \cdot A_i~\mapsto~u A_i u^{-1}+u\partial_i u^{-1}.
\end{equation}

There are two normal subgroups of $\mathcal{G}$ of importance here,
\begin{align}
  \mathcal{G}^\infty &= \big\{ u\in\mathcal{G},\quad u(x)\rightarrow~\text{the identity $e$ of $SU(3)$ as}~ |\vec{x}| \to \infty\big\}, \\
  \mathcal{G}^\infty_0 &= ~ \textrm{connected component of } \mathcal{G}^\infty.
\end{align}
As discussed elsewhere \cite{Balachandran:2013wsa,Balachandran:1991zj}, the Gauss law generates $\mathcal{G}^\infty_0$ which therefore 
acts trivially on the physical quantum states. 

The group $\mathcal{G}^\infty_0$ is normal in $\mathcal{G}^\infty$ and
\begin{equation}
  \mathcal{G}^\infty / \mathcal{G}^\infty_0 = \mathds{Z}.
\end{equation}
Its representations $\mathds{Z}\ni n~\rightarrow~e^{in\theta}$ characterise the $\theta$-states of $QCD$.

The colour group is
\begin{equation}
  \mathcal{G} / \mathcal{G}^\infty = SU(3).
\end{equation}

All observables commute with the Gauss law, that is, $\mathcal{G}^\infty_0$. In quantum physics, observables are also \emph{local} 
\cite{Haag:1992hx}, that is, they are obtained by smearing standard quantum fields with test functions with supports in compact spacetime regions. 
In the canonical formalism, that means that local observables are obtained from smeared quantum fields over a compact\footnote{Instead of a 
fixed time slice, if one considers a time average of a field $\varphi(\vec{x},t)$ over given arbitrarily small, but finite time slices, matrix elements of 
fields become smooth functions on $\mathds{R}^3$ \cite{borchers}.} spatial region $K$. Call such a field $\varphi(K)$. The action of 
$u \in\mathcal{G}$ on $\varphi(K)$ depends only on the restriction $u|_K$ of $u$ to $K$. But $u|_K$ can be smoothly extended beyond $K$ to a 
gauge transformation $u'\in\mathcal{G}^\infty_0$. There are many ways of doing so and for each $u'$, by Gauss law, if $\varphi(K)$ is an 
observable, then
\begin{equation}
  u'\varphi(K)=\varphi(K)u'.
\end{equation}
Hence,
\begin{equation}
  u\varphi(K)=\varphi(K)u,
\end{equation}
so that \emph{all local observables commute with elements of $\mathcal{G}$}.

The configuration space $Q$ for local observables is thus associated with $Q=\mathcal{A}/\mathcal{G}$ and not $\mathcal{A}/\mathcal{G}^\infty_0$ 
as naive considerations using the Gauss law would suggest.

Quantum vector states $\Psi$ instead can be built from maps from $\mathcal{A}$ to $\mathds{C}$ which are annihilated by the Gauss law:
\begin{equation}
\begin{split}
  \Psi:A &\rightarrow~\mathds{C}, \\[0.2cm]
  (u\cdot\Psi)(A) &= \Psi(A),
\end{split}
\qquad
\begin{split}
  \Psi(A)&\in\mathds{C}, \\[0.2cm]
  {\rm if}~u&\in\mathcal{G}^\infty_0.  
\end{split}
\end{equation}
Hence wave functions are sections of vector bundles built on $\mathcal{A}/\mathcal{G}^\infty_0$. It follows that we have the fibre bundle structure
\begin{equation}
  \label{fibre-bundle-1}
  \pi:\mathcal{A}/\mathcal{G}^\infty_0 ~\rightarrow~\mathcal{A}/\mathcal{G},
\end{equation}
for the group
\begin{equation}
  \label{structure-group-1}
  \mathcal{G}/\mathcal{G}^\infty_0=SU(3)\times \mathds{Z}.
\end{equation}

{\it Any function on $Q=\mathcal{A}/\mathcal{G}$ is invariant under gauge transformations, and is hence a colour singlet.}

The bundle (\ref{fibre-bundle-1}) is twisted. Otherwise we would conclude that 
$\mathcal{A}/\mathcal{G}^\infty_0=\mathcal{A}/\mathcal{G} \times \left(SU(3)\times \mathds{Z}\right)$, which is false since 
$\mathcal{A}/\mathcal{G}^\infty_0$ is connected. This last statement follows from the fact that $\mathcal{A}$ itself is connected.

This argument however must be sharpened since $\mathcal{G}/\mathcal{G}^\infty_0$ does not act freely on $\mathcal{A}/\mathcal{G}^\infty_0$. 
Indeed, an element of $\mathcal{A}/\mathcal{G}^\infty_0$ is
\begin{equation}
  \mathcal{G}^\infty_0 A := \left\langle u \cdot A, u\in\mathcal{G}^\infty_0 \right \rangle.
\end{equation}
The action of $h\mathcal{G}^\infty_0\in\mathcal{G}/\mathcal{G}^\infty_0$ on this element is
\begin{equation}
  \mathcal{G}^\infty_0 A ~\to~ (h\mathcal{G}^\infty_0) \cdot~\mathcal{G}^\infty_0 A = \mathcal{G}^\infty_0 \left(h\cdot A \right),
\end{equation}
since $\mathcal{G}^\infty_0$ is normal in $\mathcal{G}$. To see explicitly that the action is not free, choose $A=\lambda_8 a_8$ and 
$h\in SU(2) \subset SU(3)$ with 
Lie algebra basis $\lambda_1,\lambda_2,\lambda_3$ to find that $h\mathcal{G}^\infty_0$ leaves $\mathcal{G}^\infty_0 A$ invariant. Hence 
$\mathcal{G}/\mathcal{G}^\infty_0$ does not act freely on $\mathcal{A}/\mathcal{G}^\infty_0$. 

The centre $\mathds{Z}_3$ of $SU(3)$ leaves all vector potentials $A$ invariant, so we can change $\mathcal{G}$ to 
${Ad}\, \mathcal{G}=\mathcal{G}/\mathds{Z}_3$, and correspondingly define ${Ad}\, \mathcal{G}^\infty$ and  ${Ad}\, \mathcal{G}^\infty_0$.

We next consider \emph{generic} connections $\mathcal{A}_0$ with holonomy at any point $\vec{x}_0\in\mathds{R}^3$ being ${Ad}\, SU(3)$. Then 
the above ${Ad}$ groups act freely on $\mathcal{A}_0$ \cite{Singer:1978dk,Narasimhan:1979kf}, so that we obtain the principal fibre bundle
\begin{equation}
  \label{principle-bundle-2}
  \begin{split}
    \pi:\mathcal{A}_0/{Ad}\,\mathcal{G}^\infty_0 &\to~(\mathcal{A}_0/{Ad}\,\mathcal{G}^\infty_0)/\left({Ad}\,
    \mathcal{G}/ {Ad}\,\mathcal{G}^\infty_0\right), \\[0.2cm]
    {Ad}\,\mathcal{G}/{Ad}\,\mathcal{G}^\infty_0 &\simeq {Ad}\,SU(3)~\times~\mathds{Z}.    
  \end{split}
\end{equation}
Previous authors \cite{Singer:1978dk,Narasimhan:1979kf} had shown that this bundle is twisted, that is, non-trivial,
\begin{equation}
	\label{bundle-inequality-1}
  \mathcal{A}_0/{Ad}\,\mathcal{G}^\infty_0~\neq~(\mathcal{A}_0/{Ad}\,\mathcal{G}^\infty_0)/\left({Ad}\,
  \mathcal{G}/ {Ad}\,\mathcal{G}^\infty_0\right)~\times~\left({Ad}\,SU(3)~\times~\mathds{Z} \right).
\end{equation}
A quick proof is due to Singer, see his Theorem 2 in \cite{Singer:1978dk}. He starts with the fact that $\pi_j\left(\mathcal{A}_0\right)=\{0\}$, for any 
$j\in\mathds{N}$ where $\pi_j(\mathcal{A}_0)$ is the $j$th homotopy group of $\mathcal{A}_0$. In particular, since 
$\pi_0\left(\mathcal{A}_0 \right)=\{0\}$, then $\pi_0\big({\rm LHS~of~}(\ref{bundle-inequality-1})  \big) = \{0\}$. But on the RHS of 
(\ref{bundle-inequality-1}), we have that $\pi_0\big(\mathds{Z}\big)=\mathds{Z}$. Also, $\pi_1\big({\rm LHS~of~}(\ref{bundle-inequality-1})  \big) = \{0\}$, 
since $\pi_0\left({Ad}\,\mathcal{G}^\infty_0 \right)=\{0\}$, while on the RHS we have $\pi_1\left({Ad}\,SU(3) \right)=\mathds{Z}_3$. Thus, since the 
LHS and RHS of (\ref{bundle-inequality-1}) have different homotopy  groups, we conclude that they cannot be equal. For a related discussion of the relevant cohomologies, see \cite{Asorey:1986vr}.

The non-generic connections lead to some sort of boundary points. More precisely, these ``boundary points'' give a ``stratified'' manifold 
\cite{Singer:1981xw}. 

A similar situation is already known to happen in a different context. Recall the treatment of $N$ identical particles 
on $\mathds{R}^d$ \cite{Balachandran:1991zj}. In this case, the bundle space is
\begin{equation}
  \overline{Q}_N=\left\{(x_1,...,x_N),\quad x_i\in\mathds{R}^d \right\},
\end{equation}
whereas the configuration space is
\begin{equation}
  \overline{Q}_N/S_N=\left\{ [x_1,...,x_N]\right\},
\end{equation}
where $S_N$ acts by permutations of $x_i$'s and $[x_1,...,x_N]$ is an \emph{unordered} set
\begin{equation}
  [x_1,...,x_i,...,x_j,...,x_N]=[x_1,...,x_j,...,x_i,...,x_N].
\end{equation}
But if $x_i=x_j$, for some $i,j\in\{1,...,N\}$, then $(x_1,x_2,...,x_N)$ is invariant under the transformation $x_i\leftrightarrow x_j$, so that the action 
of $S_N$ on $\overline{Q}_N$ is not free. Hence to get a genuine fibre bundle, we exclude coincidence of any two points and work with
\begin{equation}
  \overline{Q}^{~0}_N=\left\{(x_1,...,x_N), \quad x_i\neq x_j \text{ if } i\neq j \right\}.
\end{equation}
Then 
\begin{equation}
  S_N: \overline{Q}^{~0}_N~\to~Q_N\coloneqq\overline{Q}^0_N/S_N
\end{equation}
gives a principle fibre bundle. This bundle is also twisted.

Given an operator like the Laplacian $\Delta$ on $\overline{Q}^{~0}_N$, the points of $\overline{Q}_N$ with $x_i=x_j$ turn up as ``boundary 
points'' where suitable boundary conditions have to be imposed.

Likewise, the non-generic connections may have to be treated by suitable conditions in an appropriate setting. They are conjectured to lead to 
different phases of QCD. We will take up these issues in another paper. But we will not encounter the need for such conditions in the approach 
taken here.

Since the bundle (\ref{fibre-bundle-1}) is twisted, previous works \cite{Singer:1978dk,Narasimhan:1979kf} infer that $SU(N)$ (or $U(N)$) gauge 
theories do not admit global gauge conditions.

In conclusion, we have the twisted bundle (\ref{fibre-bundle-1}) in QCD. Wave functions are functions on $\mathcal{A}_0/{Ad}\, \mathcal{G}^\infty_0$ 
which under ${Ad}\,SU(3)~\times~\mathds{Z}$ transform by one of its unitary irreducible representations (UIR's). Local observables instead are 
colour singlets.

\section{How Mixed States Arise}

The UIR's $n~\to~e^{in\theta}$ of $\mathds{Z}$ lead to $\theta$-states. We will remark on them in Section 8.

For now, we focus on $SU(3)$. Hence consider the wave functions
\begin{align}
\label{su3-wave-function-1}
  \big|[a_0];\rho,\lambda\big) &, \quad a_0\in\mathcal{A}_0, \\
  a_0,~a'_0\in [a_0] ~ &\Leftrightarrow ~ a'_0=u\cdot  a_0, \quad u\in\mathcal{G}^\infty_0, \nonumber  
\end{align}
transforming as the component $\lambda$ of the UIR $\rho$ of $SU(3)$
\begin{equation}
\label{su3-wave-function-2}
  \big|[u\cdot a_0];\rho,\lambda\big) = \big|[a_0];\rho,\lambda'\big)~D^\rho_{\lambda' \lambda}(u), \quad u\in SU(3).
\end{equation}
The corresponding density matrix, from which the state on the space of observables is defined, is
\begin{equation}
\label{su3-state-1}
  \omega\left([a_0];\rho,\lambda \right) = \big|[a_0];\rho,\lambda\big) \big([a_0];\rho,\lambda\big|,
\end{equation}
where we assume for simplicity that the kets are normalised to 1 in a suitable scalar product. (Actually, we must really consider wave packets in 
$[a_0]$).

The observable algebra we work with is the algebra $\mathcal{C}$ of colour singlet operators. They are associated with $\mathcal{A}/\mathcal{G}$. 
$\mathcal{C}$ contains $\mathds{1}$. We assume that it is a $C^*$-algebra, though this point does not enter the formal considerations here. 
The algebra $\mathcal{C}_\text{Loc}$, the algebra of local observables,  is a subalgebra of $\mathcal{C}$, so that a mixed state on $\mathcal{C}$ 
remains mixed when restricted to $\mathcal{C}_\text{Loc}$. In what follows, we work with $\mathcal{C}$ itself.

If $b\in\mathcal{C}$, then its mean value in the state (\ref{su3-state-1}) is
\begin{equation}
  \omega([a_0];\rho,\lambda)\big(b\big)=\big([a_0];\rho,\lambda\big|~b~\big|[a_0];\rho,\lambda\big).
\end{equation}
If $\rho$ is the colour singlet representation, the state (\ref{su3-state-1}) restricted to $\mathcal{C}$ is pure. But that is not the case if $\rho$ is a 
non-trivial $SU(3)$ UIR. We now show this result using the GNS construction. The argument is modelled on our previous work on 
ethylene \cite{Balachandran:2013kia}.

Suppose now that $\rho$ is a non-trivial $SU(3)$ UIR. We introduce the vector states
\begin{equation}
  |b\rangle, \quad b\in\mathcal{C},
\end{equation}
and the inner product
\begin{equation}
 \label{inner-product-1}
  \langle b'| b \rangle = \omega([a_0];\rho,\lambda)\left({b'}^* b \right).
\end{equation}
We emphasize that the GNS inner product $\langle \cdot | \cdot \rangle$ is different from $(\cdot|\cdot )$.

Consider the projector
\begin{equation}
  \mathds{P}=\sum_\lambda |[a_0];\rho,\lambda)([a_0];\rho,\lambda|\equiv \sum_\lambda \mathds{P}_\lambda,
\end{equation}
which is a colour singlet and hence is an element of $\mathcal{C}$. Further if
\begin{equation}
0\neq n\in\mathcal{C}~\Big( |\mathds{1}\rangle\langle \mathds{1}|-\mathds{P} \Big)  \coloneqq \mathcal{N}, \quad \text{the Gelfan'd ideal,}
\end{equation}
then $n$ is a null vector, that is,
\begin{equation}
  \langle n | n \rangle =0.
\end{equation}
Thus we introduce the equivalence classes
\begin{equation}
  \widetilde{b}=\{b+n,~n\in\mathcal{N}\},
\end{equation}
and the vector $|\widetilde{b}\rangle$, so that
\begin{equation}
  \langle \widetilde{b'}|\widetilde{b}\rangle=\omega([a_0];\rho,\lambda)(b'^* b).
\end{equation}
There are no non-zero null vectors among $|\widetilde{b}\rangle$. The completion of $\{|\widetilde{b}\rangle\}$ in the scalar product 
(\ref{inner-product-1}) gives the Hilbert space $\mathcal{H}_{\text{GNS}}$. 

The representation $\sigma$ of $\mathcal{C}$ on $\mathcal{H}_{\text{GNS}}$ is
\begin{equation}
	\label{representation-1}
  \sigma(c)|\widetilde{b}\rangle = |\widetilde{cb}\rangle.
\end{equation}
The vector $|\widetilde{\mathds{1}}\rangle$ is cyclic in $\mathcal{H}_{\text{GNS}}$, so that all of 
$\mathcal{H}_{\text{GNS}}$ can be obtained from the action of the elements of $\mathcal{C}$ (and its completion in the 
$\mathcal{H}_\text{GNS}$ norm), and
\begin{equation}
  \omega([a_0];\rho,\lambda)(c)=\Tr_{\mathcal{H}_{\text GNS}}\Big( |\widetilde{\mathds{1}}\rangle\langle
  \widetilde{\mathds{1}}|~\sigma(c) \Big).
\end{equation}

Now, the representation (\ref{representation-1}) is reducible showing that $\omega([a_0];\rho,\lambda)$ is not pure. We can see this as follows. 
Since $\mathds{1}-\mathds{P}\in\mathcal{N}$,
\begin{equation}
\label{identity-1}
  |\widetilde{\mathds{1}}\rangle=|\widetilde{\mathds{P}}\rangle.
\end{equation}
Since $\sigma(\mathcal{C})$ is an $SU(3)$-singlet, its action does not affect $\lambda$. Hence as a state,
\begin{align}
\label{decomposition-1}
  |\widetilde{\mathds{P}}\rangle\langle \widetilde{\mathds{P}}|\Big|_{\mathcal{C}} &=\sum_\lambda 
  |\widetilde{\mathds{P}_\lambda}\rangle\langle \widetilde{\mathds{P}_\lambda}|, \\
  |\widetilde{\mathds{P}_\lambda}\rangle &:= \big|[a_0];\rho,\lambda\big\rangle \big\langle[a_0];\rho,\lambda\big|.
\end{align}
On each $|\widetilde{\mathds{P}_\lambda}\rangle$, regarded as a cyclic vector, we can build a representation of 
$\mathcal{C}$:
\begin{equation}
  \sigma(c)|\widetilde{\mathds{P}_\lambda}\rangle = |\widetilde{c\mathds{P}_\lambda}\rangle.
\end{equation}
Thus $|\widetilde{\mathds{1}}\rangle\langle \widetilde{\mathds{1}}|$ restricted to $\mathcal{C}$ is a mixture of $|\rho|$ pure states ($|\rho|$ being 
the dimension of $\rho$) and is mixed for $|\rho|\neq 1$.

As discussed elsewhere \cite{Balachandran:2012pa,Balachandran:2013kia}, the decomposition (\ref{decomposition-1}) is not unique. If 
$|\mathds{P}|$ is the rank of $\mathds{P}$, $u\in U(|\mathds{P}|)$ and
\begin{equation}
  |\widetilde{\mathds{P}'_\lambda}\rangle=|\widetilde{\mathds{P}_\sigma u_{\sigma \lambda}}\rangle,
\end{equation}
then
\begin{equation}
\label{decomposition-2}
  |\widetilde{\mathds{1}}\rangle\langle \widetilde{\mathds{1}}|\Big|_{\mathcal{C}}=\sum_\lambda 
  |\widetilde{\mathds{P}'_\lambda}\rangle\langle \widetilde{\mathds{P}'_\lambda}|. 
\end{equation}
This ambiguity introduces ambiguities in entropy.

The group algebra $\mathds{C}SU(3)$ restricted to the $\rho$-representation and $\mathds{C}U(|\mathds{P}|)$ coincide. Thus the entropy 
ambiguities emerge from unobserved colour. If colour \emph{were} part of $\mathcal{C}$, the state (\ref{su3-state-1}) would remain pure.

The following point is important. Since observables are colour singlets, we can observe only $\mathds{P}$ and not $\mathds{P}_\lambda$ or 
$\mathds{P'}_\lambda$. Hence while we can prepare the vector $|\widetilde{\mathds{P}} \rangle$ by observing $\mathds{P}$, we cannot prepare 
$|\widetilde{\mathds{P}_\lambda} \rangle$ or $|\widetilde{\mathds{P'}_\lambda} \rangle$. This with (\ref{decomposition-1}) shows another way to 
understand how mixed states arise in QCD.
 
\section{The Matrix model}

\subsection{The Case of Two Colors: A Review}

The basic work leading to this model is that of Narasimhan and Ramadas \cite{Narasimhan:1979kf}. They consider the colour group $SU(2)$ and 
the spatial slice $S^3$. We remark that as for fuzzy spheres, we can recover $\mathds{R}^3$ from $S^3$ by suitable limits. 

Narasimhan and Ramadas rigorously prove that for $N=2$, the gauge bundle 
\begin{equation}
\mathcal{G}^\infty_0 \rightarrow \mathcal{A} \rightarrow \mathcal{A}/\mathcal{G}^\infty_0
\end{equation}
is twisted and does not admit a global section (that is, a gauge fixing). For proving this result, they reduce the problem to one of studying the 
special left-invariant connections
\begin{equation}
\omega =i (\Tr \tau_i u^{-1}du) M_{ij} \tau_j,
\end{equation}
where $\tau_i$ are the Pauli matrices, $u \in SU(2)$ and $M$ is a $3 \times 3$ real matrix. The connection on spatial $S^3$ is obtained by 
diffeomorphically mapping $S^3$ onto $SU(2)$ and pulling back $\omega$. The submanifold of such $\omega$ is preserved only by the global 
$SU(2)$ adjoint action
\begin{equation}
\omega \rightarrow v \omega v^{-1}, \quad v \in SU(2),
\end{equation}
or
\begin{equation}
M \rightarrow MR^T
\end{equation}
where $R$ is the $SO(3)$ image of $v$ under the homomorphism $SU(2) \rightarrow SO(3)$. The action of $SO(3)$ on the space $\mathcal{M}_0$ 
of $3\times 3$ real matrices of rank $\geq 2$ is free and leads to an $SO(3)$ fibration
\begin{equation}
SO(3) \rightarrow \mathcal{M}_0 \rightarrow \mathcal{M}_0/SO(3).
\end{equation}
From this result, they deduce that the gauge bundle is also twisted.

\subsection{The Case of Three Colors}

We now adapt the preceding discussion to $SU(3)$.  

We start with the left-invariant one-form on $SU(3)$,
\begin{equation}
	\label{Left-inv-1-form}
  \Omega=\Tr\left(\frac{\lambda_a}{2}~u^{-1}du  \right) M_{ab} \lambda_b, \quad u\in SU(3),
\end{equation}
where $M$ is a real $8\times 8$ matrix and $\Tr$ is in the fundamental representation of $SU(3)$. These $M$'s parametrize a submanifold of 
connections $\mathcal{A}$ which captures the essential topology of current interest.

In $SU(3)$, $\mathcal{\lambda}_i$, $i=1,2,3$, generate an $SU(2)\simeq S^3$ subgroup. We map spatial $S^3$ diffeomorphically to $SU(2)$, 
\begin{equation}
S^3 \ni \vec{x} \rightarrow u(\vec{x}) \in SU(2) \subset SU(3),
\end{equation}
with a distinguished point $p$ having the image $e\in SU(3)$. A convenient choice is the Skyrme ansatz \cite{Balachandran:1991zj}
\begin{equation}
u(\vec{x}) = \left(\begin{array}{cc}
					  \cos \theta(r)+i \tau_i \hat{x}_i \sin \theta(r)   &0   \\
					  0   & 1  
				\end{array}\right),  \quad \theta(0) =\pi, \quad \theta(\infty)=0, \quad \vec{x} \in \mathds{R}^3, \,\, r\equiv |\vec{x}|
\end{equation}
Although $\vec{x} \in \mathds{R}^3$, $\lim_{r \rightarrow \infty} u(\vec{x}) = \mathds{1}$, so that $u$ gives a mapping from $S^3$ to $SU(3)$.

Now, if $X_i$ are vector fields on $SU(3)$ representing $\lambda_i$ for the right action $X_i u =-u \lambda_i/2$, then 
$\left[X_i,X_j\right]=i\epsilon_{ijk} X_k$, and
\begin{equation}
  \Omega(X_i)=-M_{ib}\frac{\lambda_b}{2}.
\label{connection}
\end{equation}
Thus on identifying spatial vector fields with $iX_j$, $j=1,2,3$, one has for the vector potentials on the spatial slice,
\begin{equation}
  A_j=-iM_{jb}\frac{\lambda_b}{2}.
 \label{matrixgaugepot}
\end{equation}

Here $M$ has no spatial dependence whereas $\mathcal{G}^\infty$ acting on $A_j$ will introduce such dependence, except at identity (since 
$U(p)=e$), and will not preserve the form of $A_j$. This submanifold is thus gauge fixed with respect to $\mathcal{G}^\infty$ (Such gauge fixation is 
not possible for the space of \emph{all} $\mathcal{A}$ since $\mathcal{A}\neq (\mathcal{A}/\mathcal{G}^\infty)~\times \mathcal{G}^\infty$).

But $SU(3)$ of colour acts on $A_j$. If $h\in SU(3)$,
\begin{align}
  A_j~\to~h~ A_j~ h^{-1} \qquad \text{ or } \qquad M~\to~M({Ad}\,h)^T. 
\label{gaugetr}
\end{align}
{\it Remark:} For later use, we now show that the action (\ref{gaugetr}) is not necessarily free. This result will not be of importance in this paper.

There are four linearly independent vectors in the octet representation of $SU(3)$ which are singlets under 
hypercharge $Y \propto \lambda_8$, since
\begin{equation}
  [\lambda_8,\lambda_i]=[\lambda_8,\lambda_8]=0, \qquad i=1,2,3.
\end{equation}
These correspond to the pions $\pi_i$ and the eta meson $\eta$. Hence if the columns of $M$ are spanned by $\pi_i$ and $\eta$, then
\begin{equation}
  M\left({Ad}\,e^{i\alpha \lambda_8}\right)=M.
\end{equation}
It follows that the ${Ad}\,SU(3)$ action on $M$ is not free if its rank is $\leq~4$.

But ${Ad}\,SU(3)$ does act freely on $\mathcal{M}^0$, the space of matrices of rank $\geq~5$. We can see this as follows. Let 
$M\in\mathcal{M}^0$ and map the columns of $M$ to the $3\times 3$ $SU(3)$ Lie algebra according to
\begin{equation}
  M_{i\alpha}~\to~M_{i\alpha}\frac{\lambda_\alpha}{2}, \qquad i\in [1,2,...,8].
\end{equation}
The action $M~\to~M~({Ad}\,h)^T$ of ${Ad}\,SU(3)$ on $M$ is equivalent to its adjoint action on $\lambda_\alpha$. So we focus on the vector 
space spanned by $\lambda_\alpha$ on which $SU(3)$ acts by conjugation.

Now if an element $h\in SU(3)$ leaves $\xi_\alpha \lambda_\alpha$ and $\eta_\alpha \lambda_\alpha$ invariant under conjugation, it also leaves 
their product invariant. So the set of such vectors left invariant under $SU(3)$ conjugation forms an algebra. So does their complex linear span. Let 
$\mathcal{F}$ denotes this complex algebra. This algebra is a $*$-algebra with the $*$ defined by hermitian conjugation $h$ being unitary. It is then 
a standard result that $\mathcal{F}$ is the direct sum of full matrix algebras. As $\mathcal{F}$ acts on $\mathds{C}^3$, we can conclude that
\begin{equation}
  \mathcal{F}=\bigoplus~{Mat}_{N_i}, \qquad \sum N_i=3.
\end{equation}

We already found an algebra $\mathcal{F}'$ fixed by hypercharge, namely
\begin{equation}
  \mathcal{F}'=\left\{\left(\begin{array}{c|c} m & 0 \\ \hline 0 & c \end{array} \right) \right\},\qquad c\in\mathds{C},
\end{equation}
the $m$ being generated by $\lambda_i$ while $\mathds{C}$ can be obtained from $\lambda_3^2$ and $\lambda_8^2$.

This $\mathcal{F}'$ is maximal if its stabiliser $h$ is not a multiple of $\mathds{1}$. For the only bigger $\mathcal{F}'$ is ${Mat}_3(\mathds{C})$, and 
if $h$ commutes with all of ${Mat}_3(\mathds{C})$, then $h$ lies in the centre of $SU(3)$. Then ${Ad}\,h$ is identity.

We have thus proved that ${Ad}\,SU(3)$ acts freely on $\mathcal{M}^0$.

{\it Remark}: For $N=2$, and the gauge group ${Ad}\,SU(2)=SO(3)$, the matrix $M$ in (\ref{connection}) is $3\times3$. Narasimhan and Ramadas 
\cite{Narasimhan:1979kf} have remarked that the $SO(3)$ action
\begin{equation}
M \rightarrow M h^T, \quad h \in SO(3)
\end{equation}
is free if the rank of $M$ is larger than one. Thus $\mathcal{M}^0$ in this case are real matrices of rank 2 or 3.

\subsection{The Matrix Model Bundle is Twisted} 

Now, the dimension of $\mathcal{M}$ is $64$. The dimension of matrices of rank $4$ is $32$. Hence their codimension is also $32$. Furthermore, 
since $\mathcal{M}$ is contractible, $\pi_j(\mathcal{M})=0$ for all $j$. Hence by Remark 3 to Theorem 6.2 in Narasimhan and Ramadas \cite{Narasimhan:1979kf}, $\pi_1(\mathcal{M}^0)=0$.That is enough to show that 
\begin{equation}
  \mathcal{M}^0\neq \mathcal{M}^0/{Ad}\,SU(3)~\times~{Ad}\,SU(3)
\end{equation}
since $\pi_1 ({Ad}\,SU(3)) = \mathds{Z}_3$.

We thus conclude that the bundle
\begin{equation}
  {Ad}\, SU(3)~\to~\mathcal{M}^0~\to~\mathcal{M}^0/{Ad}\,SU(3)
\end{equation}
captures the $SU(3)$ twist of the exact theory.

Narasimhan and Ramadas in their proof of Theorem 6.2 also show that for $N=2$, the bundle 
\begin{equation}
SO(3) \rightarrow \mathcal{M}^0 \rightarrow \mathcal{M}^0/SO(3)
\end{equation}
is twisted. This result is important for us as we also consider $N=2$ explicitly in Sections 5.1 and 6.

\subsection{The Hamiltonian for \texorpdfstring{$SU(N)$}{Lg}}

Recall that the Yang-Mills action is 
\begin{equation}
S=-\frac{1}{2g^2}\int d^4x \Tr F_{\mu \nu} F^{\mu\nu}, \quad {\rm with} \quad F_{\mu\nu}=\partial_\mu A_\nu -\partial_\nu A_\mu + [A_\mu, A_\nu]
\label{YMaction}
\end{equation}
Upon rescaling $A\rightarrow gA$, we recover the form used in perturbative QCD.

From the Hamiltonian
\begin{equation}
H = \frac{1}{2}\int d^3 x \Tr \left(g^2E_i E_i - \frac{1}{g^2}F_{ij}^2 \right), \quad E_i = \text{chromoelectric field}
\label{hamiltonian}
\end{equation}
of (\ref{YMaction}), we can easily write down the Hamiltonian for the reduced matrix model, which we will do in the next section.

As the configuration space variables for the matrix model are $M_{i \alpha}$, it is natural to 
take the $\frac{d}{dt} M_{i \alpha}$ after Legendre transformation as the conjugate of $M_{i \alpha}$. In QCD, the conjugate to the connection is 
the chromoelectric field. So we identify this conjugate operator with the matrix model chromoelectric field $E_{i \alpha}$. On quantising the 
reduced model, these satisfy
\begin{equation}
[M_{i \alpha} , E_{j \beta} ]=i \delta_{ij} \delta_{\alpha \beta}.
\end{equation}

\section{Matrix Model for \texorpdfstring{$SU(N)$}{Lg} gauge theory}

In the matrix model, $A_i$ plays the role of the vector potential. From its curvature 
$d\Omega+\Omega\wedge \Omega$, we get
\begin{equation}
  \Big(d\Omega+\Omega\wedge \Omega\Big)(iX_i,iX_j)=F_{ij}=i\epsilon_{ijk} M_{k \alpha}\frac{\lambda_\alpha}{2} 
  - i f_{\alpha \beta \gamma} M_{i\alpha} M_{j\beta} \frac{\lambda_\gamma}{2},~ ~ i=1,\cdots 3, \,\, \alpha=1,\cdots N^2-1,
\label{2-form-field}
\end{equation}
where $f_{\alpha \beta \gamma}$ are $SU(N)$ structure constants.

In the reduced matrix model, the term $-(\Tr F_{ij}F_{ij})/2g^2$ plays the role of the potential $V(M)$:
\begin{eqnarray}
&&V(M) = -\frac{1}{2g^2}(\Tr F_{ij} F_{ij}) \\
&=& \frac{1}{2g^2}\left(M_{k \alpha} M_{k \alpha} - \epsilon_{ijk} f_{\alpha \beta \gamma} 
M_{i \alpha} M_{j \beta} M_{k \gamma} + \frac{1}{2} f_{\alpha_1 \beta_1 \gamma} f_{\alpha_2 \beta_2 \gamma} 
M_{i \alpha_1} M_{j \beta_1} M_{i \alpha_2} M_{j \beta_2}\right)
\end{eqnarray}

The reduced matrix model Hamiltonian is thus 
\begin{equation}
H=\frac{1}{R}\left(\frac{g^2 E_{i \alpha} E_{i \alpha}}{2} + V(M) \right)
\label{ham2}
\end{equation}
We have introduced an overall factor of $1/R$ for dimensional reasons, $R$ having the dimension of length.

Notice that in the limit $g\rightarrow0$, the potential term $V(M)$ dominates, while the kinetic term dominates in the limit $g\rightarrow \infty$. 

As a quantum operator, $H$ is thus given by 
\begin{equation}
H=-\frac{g^2}{2}\sum_{i, \alpha} \frac{\partial^2}{\partial M_{i \alpha}^2} + V(M).
\label{ham3}
\end{equation}
It  acts on the Hilbert space of functions $\psi_i$ of $M$ with scalar product
\begin{equation}
(\psi_1, \psi_2) = \int \Pi_{i, \alpha} d M_{i \alpha} \bar{\psi}_1 (M) \psi_2 (M)
\end{equation}

\noindent {\it Previous work on Related Models:}

Savvidy has suggested a matrix model for Yang-Mills quantum mechanics \cite{Savvidy:1982jk}, which has been explored by many 
researchers. However, their arguments for arriving at the matrix model differ from ours, as does their potential.

Other investigations of Yang-Mills quantum mechanics involve approximating the gauge field by several $N \times N$ (unitary or hermitian) matrices. 
The potential $V$ has interesting properties in the large $N$ limit, and several investigations have been carried out by 
\cite{O'Connor:2006wv,DelgadilloBlando:2007vx,DelgadilloBlando:2008vi,DelgadilloBlando:2012xg}. Again, these models differ from our model, in 
that our model (\ref{ham3}) is based on a single $3\times(N^2-1)$ real matrix with a kinetic energy term.

\subsection{Simplification of Potential and its Extrema: \texorpdfstring{$SU(2)$}{Lg} Case}

Let us specialise to the case of $SU(2)$ gauge theory. Then $f_{\alpha \beta \gamma} = 
\epsilon_{\alpha \beta \gamma}$. Hence
\begin{equation}
V(M) = \frac{1}{2g^2} \left(\Tr M^T M -6 \det M +\frac{1}{2} [(\Tr M^T M)^2 - \Tr M^T M M^T M] \right)
\label{potential}
\end{equation}

Let us do the singular value decomposition (SVD) of $M$: $M = R A S^T$, where $A$ is a diagonal 
matrix with non-negative entries $a_i$, and $R$ and $S$ are real orthogonal matrices. By applying extra rotations to the right of $R$ or $S$, we can assume that $a_1\geq a_2\geq a_3\geq 0$.  With this 
decomposition,
\begin{equation}
2g^2 V(M) = (a_1^2 + a_2^2 + a_3^2) - 6 a_1 a_2 a_3 + (a_1^2 a_2^2 + a_1^2 a_3^2 + a_2^2 a_3^2)
\label{potentialSVD}
\end{equation}

Note that under gauge transformations, $M \rightarrow M R^T$ (with $R \in SO(3)$), so $V(M)$ is invariant under 
gauge transformations.

The potential is zero for $M=0$, and $M=\mathds{1}$. These two are gauge-related by a large 
gauge transformation, because $u(\vec{x})$ is a winding number 1 transformation and for $M=\mathds{1}$, $A$ is the gauge transform of the zero connection by a winding number 1 transformation. 

The minima of $V$ are given by 
\begin{equation}
\frac{\partial V}{\partial a_1} = \frac{1}{g^2}\left(a_1 -3a_2 a_3 + a_1(a_2^2+a_3^2)\right) =0
\end{equation}
and similar equations from $\partial V/\partial a_2 =0, \partial V/\partial a_3 =0$. Symmetry of the equations 
under $a_i \leftrightarrow a_j$ suggests that all $a_i$ are equal at the extremum. Putting $a_1=a_2=a_3=a$ 
immediately gives $a=0, 1/2, 1$ as the extrema.

We can look at the Hessian matrix $\rm{Hess}=[\partial^2 V/\partial a_i \partial a_j]$:
\begin{equation}
\rm{Hess}=\frac{1}{g^2}\left(
\begin{array}{ccc}
 1+a_2^2+a_3^2 & 2 a_1 a_2-3 a_3 & 2a_1 a_3 -3a_2\\
 2a_1 a_2-3 a_3 & 1+a_3^2+a_1^2 & 2a_2 a_3-3a_1 \\
 2 a_1 a_3-3 a_2 & 2 a_2 a_3-3 a_1 & 1+a_1^2+a_2^2 \\
\end{array}
\right)\end{equation}

This is positive definite at $a_1=a_2=a_3=0$ 
with eigenvalues $1/g^2,1/g^2,1/g^2$. It is also positive definite at $a_1=a_2=a_3=1$ with eigenvalues $1/g^2,4/g^2,4/g^2$. Even 
though $M=0$ and $M=\mathds{1}$ are related by a (large) gauge transformation, the Hessian has a very different 
spectrum. The physical consequences of this is unclear to us.

The Hessian at $M=\mathds{1}/2$ has eigenvalues $-1/2g^2,5/2g^2,5/2g^2$. So this extremum is a saddle point. Again, we need to 
understand the physical interpretation of this saddle point.

{\it Separation of Variables in $H$:}

The quantum mechanical Hamiltonian is given by (\ref{ham3}) and (\ref{potential}). We note that for $N=2$, its separation of variables into 
radial coordinates $a_i$ and angular coordinates ($R$ and $S$) is available in previous work \cite{zickendraht,iwai}.

\section{Spectrum of the Hamiltonian}

We will work with the Hamiltonian (\ref{ham3}) and limit ourselves here to qualitative remarks and estimates about its spectrum for 
$N=2$ and $3$.  Detailed work is in progress with S. Digal. 

The potential grows quadratically in $a_i$ as $|a_i| \rightarrow \infty$, while it is smooth elsewhere. It follows immediately that the spectrum is 
gapped as required by colour confinement, and is discrete as well.

The potential resembles that of the anharmonic quartic oscillator. In the latter case, the anharmonic term is known 
to be a singular perturbation which cannot be treated using perturbation theory \cite{Loeffel:1970fe,Graffi:1990pe,Mathews:1977rf}.

We will use variational methods to estimate energy levels. We will be guided by 
 \begin{equation}
H_0 = \frac{1}{R}\left(\frac{g^2 E_{i \alpha} E_{i \alpha}}{2} + \frac{M_{i \alpha} M_{i \alpha}}{2g^2} \right)
\end{equation}
in our choice of the variational ansatz.

The eigenfunctions of $H_0$ are of the form $f(M_{i \alpha}) e^{-M_{i \alpha} M_{i \alpha}/2g^2}$, where 
$f(M_{i \alpha})$ are products of Hermite polynomials in $3(N^2-1)$ variables $M_{i \alpha}$.

For the variational ansatz for the ground state, we take 
\begin{equation}
\Psi^0_b = A_0 e^{-\frac{b}{2g^2} M_{i \alpha} M_{i \alpha}}, \quad A_0 = \left(\frac{b}{\pi g^2}\right)^{\frac{3}{4}(N^2-1)},
\end{equation}
and minimise with respect to the parameter $b$.

We find
\begin{equation}
\langle \Psi^0_b |H |\Psi^0_b \rangle \equiv E^0(b,g)=\frac{3}{4R}(N^2-1)\left(b+\frac{1}{b}+\frac{g^2N}{2b^2}\right)
\end{equation}
Minimizing with respect to $b$ gives the variational ground state energy $E^{0}_\text{min}(g)$. It is plotted in Figure 1 as a function of t'Hooft 
coupling $t=g^2 N$.

\begin{figure}[h]
        \centerline{
               \mbox{\includegraphics*[width=4in]{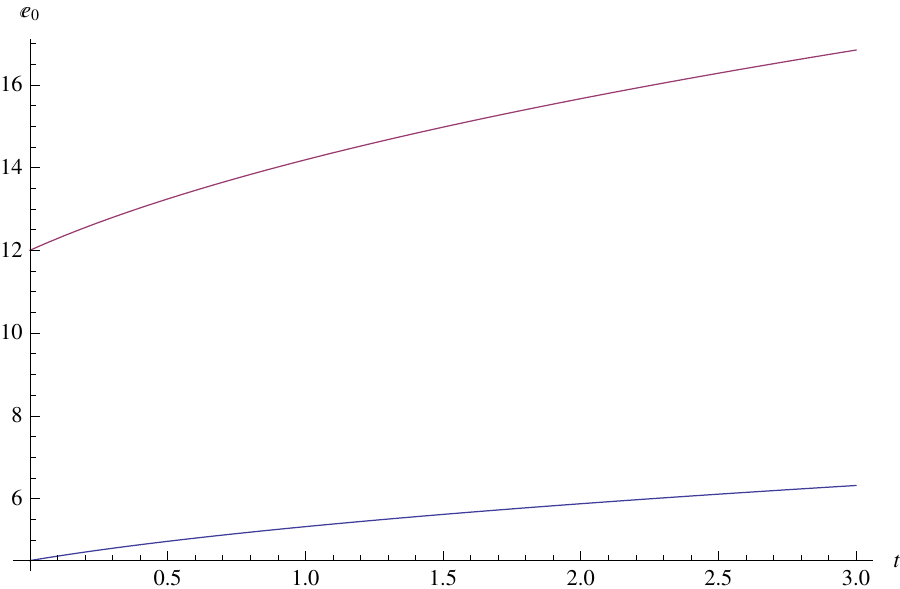}}
                             }
\caption{The blue line is for $N=2$, the red line is for $N=3$. The energy is in units of $1/R$ and $t=g^2N$.}
\end{figure}

Similarly, we can take the ansatz
\begin{equation}
\Psi^1_b = A_1 M_{i \alpha} e^{-\frac{b}{2g^2} M_{i \alpha} M_{i \alpha}}, \quad A_1 = \frac{2b}{g^2} \left(\frac{b}{\pi g^2}\right)^{\frac{3}{4}(N^2-1)},
\end{equation}
for the first excited state. This is an impure state because the colour index $\alpha$ is not "soaked up". We then calculate 
\begin{equation}
\langle \Psi^1_b |H | \Psi^1_b \rangle \equiv E^1(b,g),
\end{equation}
to find 
\begin{equation}
E^1(b,g) = \frac{1}{4R} (3 N^2 - 1) \left(b + \frac{1}{b} + \frac{g^2N}{2 b^2}\right) + \frac{g^2N}{4 b^2}.
\end{equation}

Its minimum $E^1_\text{min}(g)$ is plotted against $g^2 N$ in Figure 2.

\begin{figure}[h]
        \centerline{
               \mbox{\includegraphics*[angle=0,width=4in]{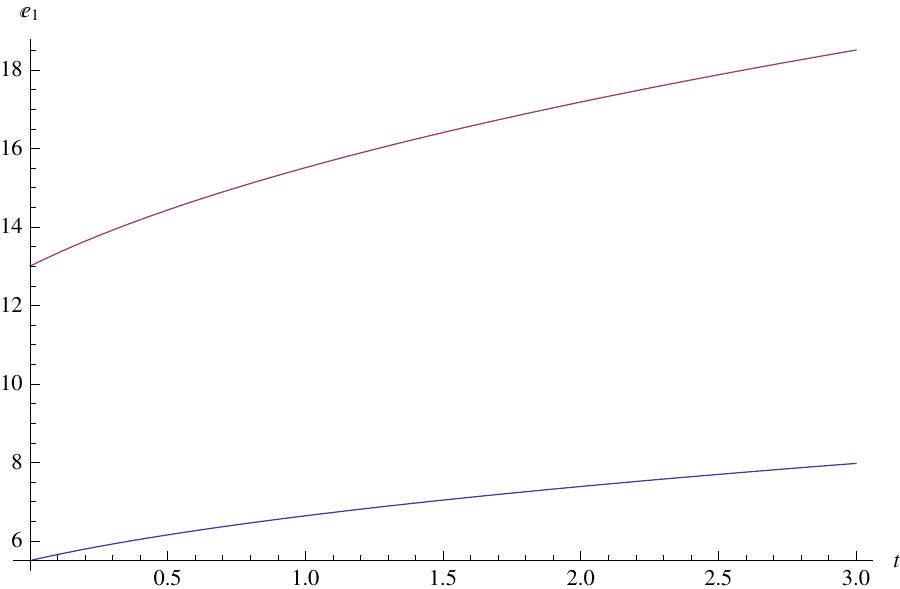}}
                             }
\caption{The blue line is for $N=2$, the red line is for $N=3$. The energy is in units of $1/R$ and $t=g^2N$.}
\end{figure}

Notice that both these trial wave functions are insensitive to the $O(g)$ term in the Hamiltonian. The simplest ansatz that is sensitive to this term is 
\begin{equation}
\Phi^1_{(b,c)} = B_1 (M_{i \alpha} +c \,\epsilon_{ijk} f_{\alpha \beta \gamma} M_{j \beta} M_{k \gamma}) e^{-\frac{b}{2g^2} M_{i \alpha} M_{i \alpha}}, \quad c \in \mathds{C}.
\end{equation}
This has three variational parameters: $c,c^*$ and $b$. The variational energy for this ansatz is shown in Figure 3.

\begin{figure}[h]
        \centerline{
               \mbox{\includegraphics*[angle=0,width=4in]{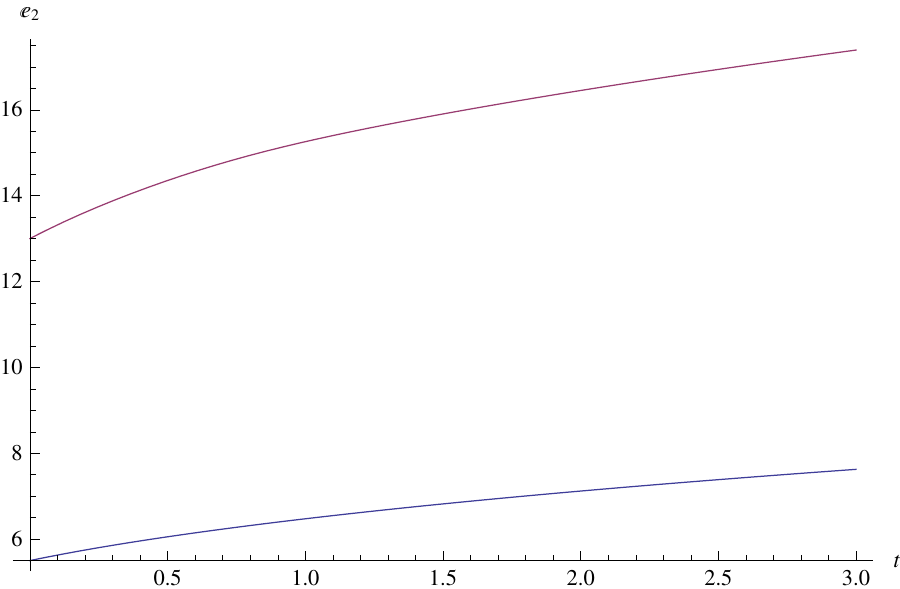}}
                             }
\caption{The blue line is for $N=2$, the red line is for $N=3$. The energy is in units of $1/R$ and $t=g^2N$.}
\end{figure}

Our variational energy estimate is rather crude, and is presented here for representational purposes only. We expect that the variational 
estimate differs significantly from the true energy  for large values of t'Hooft coupling $t$. Much better numerical estimates may be obtained by 
taking more sophisticated (or complicated!) variational ansatz for the wavefunctions. We will not do it here.

\section{On Mixed States in the Matrix Model}

Considerations using $\mathcal{A}$ in sections 2 and 3 were formal, whereas the matrix model for $N=3$ is that of 
a particle with $64$ degrees of freedom. It is a well-defined quantum mechanical model, which captures the colour 
twist topology of $QCD$.

The $C^*$-algebra $\mathcal{C}(\mathcal{M})$ of the observables are made up of colour singlets. It contains colour singlet 
functions of $M$. (More precisely, we consider only bounded operators of this sort). The full 
$\mathcal{C}(\mathcal{M})$ is generated by such operators. 

We can now adapt section 3  to show that coloured states restricted to $\mathcal{C}(M)$ are not pure.

\section{Final Remarks}

The one definite result we have in the work is the conclusion that coloured states in QCD are mixed. That will affect correlators and partition 
functions and hence physical predictions. Calculations in this directions have not been done.

In addition, we have developed a matrix model for pure QCD which gives a gapped spectrum and discrete levels for glueballs. 

Our present work can be generalised to other gauge groups.

We conclude with a few further remarks on the matrix model. 


\begin{enumerate}
    
  \item We can couple quarks to $A_i$ by using covariant derivative $\nabla_i=\partial_i+A_i$ in the Dirac 
  operators, this being its only modification in the $A_0=0$ gauge.
  
  \item We can construct QCD $\theta$-states as follows. The Chern-Simons $3$-form  gives the field theory action
  \begin{eqnarray}
    S_{\rm CS}(A)&=&\frac{1}{8\pi^2} \int\Tr\left(A\wedge F - \frac{1}{3}A\wedge A\wedge A \right), \\
    &=&\frac{1}{16\pi^2} \int d^3 x \epsilon^{ijk} \Tr \left(A_i(x) F_{jk}(x)-\frac{2}{3} A_i(x) A_j(x) A_k(x) \right)
  \end{eqnarray}
  which in the matrix model becomes, on using (\ref{matrixgaugepot}) and (\ref{2-form-field}),
  \begin{equation}
    S_{\rm CS}(M)=\frac{1}{4} \Big[ \Tr(M^T M) +\frac{1}{6} \epsilon_{ijk} f_{\alpha\beta\gamma} M_{i\alpha}M_{j\beta}M_{k\gamma}\Big].
  \end{equation}
  The overall $1/4$ is fixed by requiring that for a pure gauge, where $M=\bf{1}_{3\times 3} \oplus \bf{0}
  _{5\times 5}$, where $\bf{1}_{3\times 3}$ is in the $SU(2)$ subspace, the RHS becomes the winding number $1$. Then under 
  a gauge transformation
  \begin{equation}
    A~\to~hAh^{-1}+hdh^{-1},
  \end{equation}
  $S_{\rm CS}(A)$ changes by the winding number $N(h)$ of the map $h$ \cite{Balachandran:1991zj,hoppe}:
  \begin{equation}
    S_{\rm CS}(hAh^{-1}+hdh^{-1})=N(h)+S_{\rm CS}(A), \quad N(h) \in \mathds{Z}.    
  \end{equation}
Hence
\begin{equation}
  e^{i\theta S_{\rm CS}(hAh^{-1}+hdh^{-1})}=e^{i\theta N(h)}e^{i\theta S_{\rm CS}(A)}.
\end{equation}

Thus given a vector state $|\cdot,\theta=0\rangle$ for $\theta=0$, we can get the one $|\cdot, \theta \rangle$ with non-zero $\theta$ as follows:
\begin{equation}
|\cdot,\theta\rangle=e^{i\theta S_{\rm CS}(M)}~|\cdot,\theta=0\rangle.
\end{equation}
With this formula, concrete calculations can be done using the Hamiltonian $H$.  

\item We can build multiparticle states for our gluon levels from (\ref{Left-inv-1-form}) by changing $u(\vec{x})$ to higher winding number maps as in Skyrmion physics \cite{Balachandran:1991zj}.

\item That coloured states are impure states have deep implications for the confinement problem. Consider the time-evolution of a pure (and 
hence colourless) state. Since time evolution in quantum theory is given by a unitary operator, this state will {\it never} evolve to a coloured 
state. Thus it is impossible to create a free gluon starting from a colourless state by any Hamiltonian evolution, and in particular by scattering. 

\item The mixed coloured states we obtain are convex combinations of pure states. Of these, at most one can be in the domain of the Hamiltonian. 
 Consequently, the mean energy in such mixed states is infinite \cite{Balachandran:1983fg,Balachandran:2011bv,Gupta:2013fwa}. This is an 
additional argument in support of the relevance of these mixed states for the confinement problem in Yang-Mills theories.

\end{enumerate}

\section{Acknowledgements}
We are very grateful to M. S. Narasimhan for many discussions and inputs. We have also benefitted from the suggestions of Sanatan Digal, Denjoe O'Connor and Apoorva Patel. Manolo Asorey and Juan Manuel Perez-Pardo have explained to us specific issues connected to domains of operators. APB thanks the group at the Centre for High Energy Physics, IISc, Bangalore, and especially Sachin Vaidya for hospitality. AQ thanks the DFT of the Universidad de Zaragoza for the hospitality and nice atmosphere. In particular, AQ thanks Monolo Asorey for fruitful discussions.  AQ is supported by CAPES process number BEX 8713/13-8.

\end{document}